\documentclass[letterpaper,twocolumn,10pt]{article}
\usepackage{SOUPS}

\usepackage{amsmath}

\usepackage{todonotes}
\usepackage{tabularx}
\usepackage{hyperref}
\usepackage{xspace}
\usepackage{url}
\usepackage{float}
\usepackage{pdflscape}
\usepackage{booktabs}
\usepackage{siunitx}
\usepackage{adjustbox}
\usepackage{array}

\newcolumntype{R}[2]{%
    >{\adjustbox{angle=#1,lap=\width-(#2)}\bgroup}%
    l%
    <{\egroup}%
}
\newcommand*\rot{\multicolumn{1}{R{45}{1em}}}%
\usepackage{tikz}
\newcommand*{\present}[1]{\begin{tikzpicture}[scale=0.15]%
    \draw (0,0) circle (1);
    \fill[fill opacity=1,fill=black] (0,0) -- (90:1) arc (90:90-#1*3.6:1) -- cycle;
    \end{tikzpicture}}

\newcommand{\reddit}{Reddit\xspace}

\newcommand{\survp}{SP\xspace}
\usepackage{chngpage}

\usepackage{filecontents}

\begin{document}

\date{}

\title{\Large \bf Peer Surveillance in Online Communities}

\def\plainauthor{Kyle Beadle and Marie Vasek}

\author{
{\rm Kyle Beadle and Marie Vasek}\\
University College London
} 

\maketitle
\begin{abstract}
Online communities are not safe spaces for user privacy.
Even though existing research focuses on creating and improving various content moderation strategies and privacy preserving technologies, platforms hosting online communities support features allowing users to surveil one another--leading to harassment, personal data breaches, and offline harm.
To tackle this problem, we introduce a new, work-in-progress framework for analyzing data privacy within vulnerable, identity-based online communities.
Where current SOUPS papers study surveillance and longitudinal user data as two distinct challenges to user privacy, more work needs to be done in exploring the sites where surveillance and historical user data assemble.
By synthesizing over 40 years of developments in the analysis of surveillance, we derive properties of online communities that enable the abuse of user data by fellow community members and suggest key steps to improving security for vulnerable users.
Deploying this new framework on new and existing platforms will ensure that online communities are privacy-conscious and designed more inclusively.
\end{abstract}
\thecopyright

\section{Introduction}
\label{sec:intro}

In 2012, LGBTQ+ communities erupted on the social networking site, \reddit.
Disagreements over content moderation led to many users leaving r/lgbt for the newly-founded r/ainbow~\cite{gibson:sms2019}.
The creation of more niche groups followed as users of different identities felt under-served by broader groups.
For example, users on r/transgender and r/bisexual were opposed by other community members for creating posts that strayed from the typical submissions in those communities -- leading to the formation of r/transtimelines, r/nonbinary, and more.
These splintering events resulted in new discussions, new norms, and new community identities for users.

Everyday users generate millions of data points when they post on online communities such as \reddit.
Unless a community is private or age-gated, all of this data is visible to anyone, anywhere, regardless of whether they use an account or not.
As a result, the comments and submissions of particular users within online communities can be observed over time and analyzed.
If an offline identity can be linked to these posts, a person could be doxxed\footnote{The online act of disclosing identifying or sensitive information with the malicious intent to harm another~\cite{snyder:imc2017}. This is often against platforms' Terms of Service.} based on their posts.
Users also follow and harass one another from community-to-community on the basis of previous activity.

When we refer to an online community, we speak specifically of identity-based communities -- the terms will be used interchangeably hereafter.
An identity-based community is a collective of individuals that share circumstances instead of geographical location~\cite{mcmillan:jcp1986}. 
Online communities do create identities around video games played, films watched, or politics discussed, yet our focus lies in marginal identities that organize online~\cite{hwang:cscw2021}.
These communities centered on shared demographics create environments of solidarity and support~\cite{tausczik:cscw2014}. 

We argue that peer surveillance, as opposed to top-down surveillance, provides a new perspective with which to re-examine how members of online communities interact with one another.
Where top-down surveillance is an authoritarian act of the more powerful observer monitoring the less powerful observed, peer surveillance is much more nuanced.
Peer surveillance is the act of using observation to influence the behavior of friends, family, and colleagues. 
We, therefore, propose a research agenda focused on exploring how online, identity-based communities deploy peer surveillance to shape digital behaviour and censor users.
By recognizing the diversity and fluidity of online communities, our work can further be applied to frame user studies of privacy within individual communities and to the development of inclusive privacy features for platforms that host identity-based communities.
\section{Related Work}
\label{sec:background} 

Other work by the SOUPS community recognize the influence of surveillance and the privacy challenges of longitudinal data on social networking sites.
For example, surveillance and social influence is experienced by children whose caretakers install parental control apps on their devices~\cite{ghosh:chi2018} and is questioned by the relatives of inmates during phone calls~\cite{owens:chi2021}.
These studies focus on the large power differential between two groups, parents and children and prison guards and inmates, rather than the slight differences between coworkers or fellow community members.
Keeping online data visible for a long time additionally challenges user privacy~\cite{mohamed:soups18,mazzia:soups2012,ayalon:soups2013,strater:soups2007}.

Yet, the properties of online platforms which enable peer surveillance of vulnerable users remain understudied by the SOUPS community.
Das et al.~\cite{das:soups2014} highlight that people's security behaviors are influenced by others while Rashidi et al.~\cite{rashidi:soups2018} present how college students manage their privacy when they are constantly surveilled by classmates.
Peer surveillance impacts vulnerable groups differently than the general population.
The result of this gap is online communities with security and privacy features that are hostile to the vulnerable groups hosted there.

Opening research into peer surveillance in online communities will protect the under-served populations that Wang~\cite{wang:nspw2017} argues in favor of including in usable security and privacy.
All communities do require their own individual needs and inspections, but many communities are hosted on similar platforms which do consider their most vulnerable users -- whether it be the visually-impaired~\cite{ahmed:chi2015,hayes:soups2019}, older adults~\cite{jovanovic:hci2021}, Black Lives Matter protesters~\cite{wade:hci2021}, trans activists~\cite{jackson:nms2018}, Asian American and Pacific Islanders~\cite{dosono:hci2019}, veterans~\cite{zhou:cscw2022}, or individuals struggling with mental health~\cite{sharma:chi2018}. 
Focusing on more content moderation, which is imposed onto communities, ignores the ways in which community members abuse platform features and user data to target other community members.
Online communities require more usable privacy features to combat malicious members who skirt content policies to coerce fellow individual members.
\section{Surveillance}
\label{sec:surveillance}

Surveillance studies is a multidisciplinary project, spanning political science, sociology, law, and criminology, that describes the processes and outcomes of motivated observation.
In 1977's \textit{Discipline and Punishment}, Foucault linked the act of observation with the exercise of power.
Deploying the conceptual panopticon, he described how the threat and uncertainty of being watched leads to compliance among the observed~\cite{foucault:1977}.
Surveillance in the panopticon is top-down and allows the few, and the more powerful, to observe the many.
Recent work in surveillance studies contends that modern surveillance goes beyond the panopticon and proliferates through all areas of modern life~\cite{haggerty:2006}.

\subsection{Introducing Surveillance Studies}

In introducing surveillance studies, we present the recent developments and principles that guide a surveillance analysis of data collection practices.

\paragraph{Principle 1: Surveillance is the everyday local, global, and routine collection of personal information for analysis~\cite{lyon:ics2002}.}
\label{p:sp1}

Analysis of surveillance practices seek to prove the occurrence of surveillance through all parts of daily life.
When surveillance is not confined by direct human observation, it is not just online fandom facilitating Chinese authoritarianism~\cite{luo:nms2022}; it is unintentionally identifying faceless photographs on \reddit~\cite{van:mi2020} and other indirect acts.
These \textit{unfocused interactions} persist as surveillance because even though creating a record is not the original intent of the observer, any record can be created post-observation~\cite{eley:ss2020}.

Additionally, surveillance is not possible unless it is routine.
New users of Slashdot regularly observe other members to learn acceptable behavior before posting~\cite{lampe:group2005}.
And the daily act of reading and engaging in fan fiction results in experienced writers mentoring new participants; informal learning through surveillance uplifts the writing of the mentee to community standards~\cite{campbell:cscw2016}.

\paragraph{Principle 2: The prevalence of communication technology facilitates everyday surveillance~\cite{andrejevic:2004,mathiesen:tc1997,lyon:bds2014}.}
\label{p:sp2}

One critique of Foucault, by Mathiesen, was that Foucault did not consider the impact of technology on surveillance~\cite{mathiesen:tc1997}. 
Television, the internet, and other new media technologies from the second half of the twentieth century allowed for what Mathiesen called the \textit{synopticon}, where the many, and less powerful, now observe the few.
For example, being able to view which content is removed enables online, pro-eating disorder communities to circumvent content moderation~\cite{gerrard:nms2018}.
Communication technology also changes the intent behind deploying surveillance, constructing surveillance as an anticipatory act.
Automatic moderation techniques, such as in Chandrasekharan et al., use historical data from online communities to then moderate those communities, establishing a use of data to be applied to problems beyond the goals of the initial posting.~\cite{chandrasekharan:cscw2019}

\paragraph{Principle 3: Computers facilitate the classification of personal information into social and economic groups with the goal of influencing and managing individuals and populations~\cite{browne:2015,lyon:ss2005}.}
\label{p:sp3}

Data retrieved from surveillance technology does not exist without purpose.
Surveillance distinguishes individuals from one another, yet the development of computers changed the reach and speed at which surveillance categorization occurs~\cite{lyon:ss2005}.

Uneven outcomes for different groups~\cite{gandy:p1993} and the reification of pre-existing classifications~\cite{poudrier:ss2005} result from this computational classification.
For example, moderators on the video live-streaming website, Twitch, directly observe livestreams and deploy a series of categorization techniques to remove undesired comments~\cite{cai:hci2022}.
These techniques are often algorithmic and inherit the biases of the manually moderated content that it is trained on~\cite{binns:socinfo2017}.
On \reddit, an automatic moderation tool, Automod, returns varying levels of false positives depending on the community~\cite{jhaver:hci2019}.
As a result of the speed computers bring, the justifications and subjects of surveillance rapidly change to reflect the changes in political, social, and behavioral conditions.

\paragraph{Principle 4: Surveillance use-cases are not always predetermined; they also emerge from new applications of systems created for very different purposes~\cite{haggerty:2006}.}
\label{p:sp4}

The advent of social media presented new opportunities to the analysis of surveillance as people became more willing than ever before to share personal data online.
For example, League of Legends players habitually flag other players for ineffective teamwork when the intended use of flagging is to stop toxic behavior~\cite{kou:chi2021}.
Members of the online community r/RoastMe invite biting, comedic judgments of users despite Reddit's safety rules at large.
In both cases, the act of flagging users and commenting on their appearance flips the platform features intended usage into a surveillance act intended to change another user's behavior.

\paragraph{Principle 5: Surveillance appears in different forms, dynamics, and societal levels~\cite{andrejevic:ss2002,haggerty:scs2017,bauman:2005}.}
\label{p:sp5}

With the internet and social media, surveillance requires no territory and mutates as different motivations arise.
Instead, surveillance systems allow data to be pulled away from their initial context and rebuilt under different circumstances~\cite{haggerty:scs2017}. 
This means that surveillance is not restricted to one location, such as using online communities to trace COVID-19 in North Carolina~\cite{whitfield:bcbhi2021}.
Nor is one specific person or group relegated to being the observer or the observed.
Instead, anyone, anywhere can conduct surveillance against everyone, everywhere, such as community members ensuring that other members remain sober~\cite{gauthier:hci2022} or harassment of women and LGBT live streamers from people they do not know~\cite{uttarapong:icime2021}. 
These two examples show how social media expands surveillance beyond top-down, localized relationships to enable multiple, global dynamics that exist simultaneously.

The idea of \textit{peer surveillance} also emerged from social media.
Andrejevic defines peer surveillance as "the need to enlist monitoring strategies as a means of taking responsibility for one’s own security in a networked communication environment in which people are not always what they seem"~\cite{andrejevic:ss2002}.
Unlike top-down surveillance, where a clear power relationship exists between the more powerful observer and the less powerful observed, power dynamics are more nuanced, if not equal, under peer surveillance.
Online, anyone with an internet connection can log into online platforms, create an account, and closely follow anyone they desire~\cite{marwick:ss2012}.
These power relations become problematic in online communities where members perceive each other as equals.
Even though, on the surface, all members of a community might enjoy the same TV show, video game, or movie, each community member has their own internal interest in the community.
When these different ideas clash, community dynamics change as conflict emerges~\cite{billings:hci2010}.
\section{Framework and Design Proposal}

Vulnerable users engage in online communities because they provide a safety and solidarity that does not exist offline.
Yet, in practice, the community decides what is acceptable.
These behaviors are peer surveillance because any member of an online community can closely observe one another with the motivation of changing another user's behavior.
From the principles of surveillance studies presented above, we derive the existing peer-surveillance properties of online communities.
We present examples of where these properties are present in the Appendix, Table~\ref{tab:platforms}.
We then propose some steps to counter peer surveillance and to ensure inclusive privacy for online communities.

\subsection{Properties of Online Communities Enabling Peer Surveillance}
\paragraph{Unique Pseudonyms}
When joining an online community, users choose a pseudonym that link all of their community activity together. 
        In some cases, anyone can view a user's entire community activity on that user's profile page.
        Unique pseudonyms, while removing a person's real name from their online activity, allow malicious users to check-in on the activity of other users to ensure compliance -- fulfilling the requirement of routine collection laid out in \hyperref[p:sp1]{\survp 1}. 
\paragraph{Topic Segmentation} Many online communities are organized around different discussion categories.
        For example, members of a math learning forum will choose whether their question fits in the calculus, geometry, or arithmetic categories before posting~\cite{van:2009}.
        These different groupings act as silos to contain discussion and manage information overload.
        Yet, anyone can be pulled into a conversation at anytime.
        A member that posts solely in arithmetic can be tagged in the discussion of a calculus problem.
        This may be harmless in the world of mathematics but can be triggering for someone who posts in an entertainment discussion to be unexpectedly tagged in a stranger's abuse story on a forum dedicated to supporting survivors of intimate partner violence.
        As such, topic segmentation realizes \hyperref[p:sp4]{\survp 4} when paired with tagging because it enables abuse of data which was created for a very different purpose.
\paragraph{Focused Search} Online communities allow for data only within those communities to be searchable.
        This feature is obvious in internet forums where only the community exists but is also present on larger platforms that host online communities, such as Reddit, Facebook, and Discord.
        While context-based search eases information finding, it is also abused to investigate community members' prior events and to examine who commonly appears in the same discussions.
        Communication technology allows quickly searching for data across the web to be possible and online communities allow data search within more specific contexts to be possible -- satisfying \hyperref[p:sp2]{\survp 2}.
\paragraph{Trust-Based Disclosures} Users are more likely to disclose personal information in online communities because it is where they feel safe~\cite{ammari:cscw2016}. 
        However, user data remains publicly visible despite members feeling that a community is private.
        Certain personal data, such as photos, location names, and personal histories can intentionally or unintentionally connect a user’s online presence to their offline identity.
        As such, trust-based disclosures within online communities fulfill \hyperref[p:sp4]{\survp 4} because information given in confidence can be abused by other community members since it is likely that data remains visible into the future.
\paragraph{Reputation Systems} Feedback mechanisms in online communities allow users to discern whether another user is trusty-worthy or likely to misbehave~\cite{gurtler:pets2021}.
        Built-in features, such as likes, upvotes, and awards, reward members that post according to community expectations.
        Other, more social systems, such as community norms and peer pressure through private messages, steer members' into conformity.
        Both built-in and hidden, social systems highlight how platforms implement features that enable in-group coercion.
        Reputation systems meet \hyperref[p:sp3]{\survp 3} because they grant users with the knowledge to sort users on the frequency of their post activity and how other community members perceive them.
        These systems also meet \hyperref[p:sp2]{\survp 2} because community members use reputation to anticipate another member's future activity.

\subsection{Peer Surveillance-Resistant Design}
\paragraph{User Discretion} In designing more inclusive and secure online communities, user preferences should be examined and considered.
        For example, users should be able to approve when their pseudonym gets tagged into a discussion or be able to determine, upon creating their account, when they want their pseudonym to be attached to a post.
        Presenting users with more options could be overwhelming so more consideration is required.
\paragraph{Automatic Hiding} One way to avoid focused searches and abuses of trust-based disclosures would be to exclude user data from begin found after a certain amount of time.
        Additionally, all posts should eventually hide user pseudonyms from being seen as well as user posts being searched.
        Given instances where personal experiences become more vulnerable over time, such as during gender transitions or progressive illnesses, community members should have more ways to detach from years old, personal data.
\paragraph{Temproary Pseudonyms} Even though many online communities employ the use of pseudonyms, linking all personal data to a single account can lead to offline identification~\cite{bergstrom:fm2011}.
        Single-use identifiers should be explored for users wishing to whisteblow, organize, or share personal experiences.
        It is important to be consider how this feature might be abused, yet allowing moderators and administrators to see the link between account usernames and single-use identifiers could mitigate harm.
        Yet it is time to recognize that universal pseudonyms create potentially traceable data profiles.
\section{Conclusion}
Users do not have complete privacy in online communities.
Investigations into popular online communities, such as on Reddit, Twitch, and Internet forums, reveal that community-based moderators deploy surveillance tactics and abuse user data to shape the social behavior of community members, not just enforce terms of service.
Yet more work needs to be done. 
Platforms hosting online communities support features allowing users to surveil one another beyond moderation -- leading to harassment, personal data breaches, and offline harm.

By synthesizing over 40 years of developments in the analysis of surveillance, we concluded that the prevalence of communication technology facilitates the everyday, routine collection of data for the classification of users at different forms, dynamics, and societal levels.
Future work should apply these principles in qualitative studies and design proposals to consider the underlying power dynamics involved in data privacy.
More specifically, we highlight the five properties of online communities which facilitate peer surveillance; unique pseudonyms, topic segmentation, focused search, trust-based disclosures, and reputation systems.
In suggesting some keys steps to improve the privacy of online communities, we propose user discretion, automatic hiding, and single-use pseudonyms.
Deploying this new framework on new and existing platforms will ensure that online communities are privacy-conscious and designed more inclusively. 
\section{Acknowledgments}

This project was funded by the UK EPSRC grant EP/S022503/1.

\bibliographystyle{plain}
\bibliography{references}

\section{Appendix}

\begin{table}[h]
\centering
\label{labelhere}
\begin{tabular}{@{}lllllllll@{}}
\toprule
\textit{Platform}  & \textit{Name} & \rot{\textit{Unique Psuedonyms}} & \rot{\textit{Topic Segmentation}} & \rot{\textit{Focused Search}} & \rot{\textit{Trust-Based Disclosures}} & \rot{\textit{Reputation Systems}} & &  \\ \cmidrule{1-2} \cmidrule{3-8}
\multicolumn{2}{c}{\textbf{Communities}} & \multicolumn{5}{c}{\textbf{Properties}} & \multicolumn{2}{c}{\textbf{Related Work}} \\
\midrule
Discord   & Servers        &  \present{100}     &    \present{100}               &       \present{100}         &      \present{100}                   &   \present{50}    &             &     ~\cite{kiene:cscw2019}        \\
Douban    & Groups          &        \present{100}         &    \present{100}                &      \present{100}          &    \present{100}                     &       \present{100}             &   & ~\cite{luo:nms2022}\\
Facebook  & Groups          &    \present{0}         &  \present{0}    &       \present{100}             &       \present{100}         &       \present{50}                  &                    &       ~\cite{ammari:cscw2016}       \\
Reddit    & Subreddits           &  \present{100}    &    \present{100}    &    \present{100}                &    \present{100}            &    \present{100}                     &                    &      ~\cite{bergstrom:fm2011,chandrasekharan:hci2018,chandrasekharan:hci2022,dosono:cscw2020,gauthier:hci2022,gibson:sms2019,hwang:cscw2021,whitfield:bcbhi2021,sharma:chi2018}        \\
Twitch    & Channels            &    \present{100}           &    \present{100}                &   \present{0}             &      \present{100}                   &  \present{50}    &              &    ~\cite{cai:hci2022,uttarapong:icime2021}          \\
Wikipedia & The Community       &     \present{100}      & \present{100}   &     \present{0}               &     \present{0}           &     \present{100}                    &                    &    ~\cite{billings:hci2010}          \\
XenForo   & Forums              &   \present{100}      &  \present{100}    &         \present{100}           &       \present{100}         &   \present{100}                      &                    &        ~\cite{van:2009}      \\ \bottomrule
\multicolumn{3}{@{}p{1.5in}}{\footnotesize \present{100} $=$ Fully present}\\
 \multicolumn{3}{@{}p{1.5in}}{\footnotesize \present{50} $=$ Partially present}\\
\end{tabular}
 \caption{Peer Surveillance Properties of Online Communities. \\\\ We used the sites included in our references as a snapshot of current online communities and platforms. The list is not exhaustive. Even though Discord, Facebook, and Twitch have designed features for reputation, such as badges, likes, and reactions, the outcome of these features are not aggregated. Instead, reputation relies on human memory. Additionally, Discord pivoted to different username systems throughout the years. They are currently phasing out their four-digit discriminators in favor of unique alphanumeric user and a non-unique display name~\cite{usernames:discord2023}.}
 \label{tab:platforms}
\end{table}

\end{document}